\documentclass[twocolumn,superscriptaddress,aps,showpacs,amsmath,footinbib,bibnotes]{revtex4-1}
\pdfoutput=1
\usepackage{amsmath}
\usepackage{amssymb}
\usepackage{stmaryrd}
\usepackage{mathrsfs}
\usepackage[utf8]{inputenc}
\usepackage[export]{adjustbox}
\usepackage[dvipsnames]{xcolor}
\usepackage{fancyhdr}
\begin{document}

\author{Amirhassan Shams-Ansari}
\affiliation{John A. Paulson School of Engineering and Applied Sciences, Harvard University, Cambridge, Massachusetts 02138, USA}
\author{Dylan Renaud}
\affiliation{John A. Paulson School of Engineering and Applied Sciences, Harvard University, Cambridge, Massachusetts 02138, USA}
\author{Rebecca Cheng}
\affiliation{John A. Paulson School of Engineering and Applied Sciences, Harvard University, Cambridge, Massachusetts 02138, USA}
\author{Linbo Shao}
\affiliation{John A. Paulson School of Engineering and Applied Sciences, Harvard University, Cambridge, Massachusetts 02138, USA}
\author{Lingyan He}
\affiliation{HyperLight, 501 Massachusetts Avenue, Cambridge, MA 02139}
\author{Di Zhu}
\affiliation{John A. Paulson School of Engineering and Applied Sciences, Harvard University, Cambridge, Massachusetts 02138, USA}
\author{Mengjie Yu}
\affiliation{John A. Paulson School of Engineering and Applied Sciences, Harvard University, Cambridge, Massachusetts 02138, USA}

\author{Hannah R. Grant}
\affiliation{Freedom Photonics, 41 Aero Camino, Goleta CA, USA}

\author{Leif Johansson}
\affiliation{Freedom Photonics, 41 Aero Camino, Goleta CA, USA}

\author{Mian Zhang}
\affiliation{HyperLight, 501 Massachusetts Avenue, Cambridge, MA 02139}
\author{Marko Lon\v{c}ar}
\affiliation{John A. Paulson School of Engineering and Applied Sciences, Harvard University, Cambridge, Massachusetts 02138, USA}
\date{\today}

\title{Electrically pumped high power laser transmitter integrated on thin-film lithium niobate}

\begin{abstract}
Integrated thin-film lithium niobate (TFLN) photonics has emerged as a promising platform for realization of high-performance chip-scale optical systems. Of particular importance are TFLN electro-optic modulators featuring high-linearity, low driving voltage and low-propagation loss. However, fully integrated system requires integration of high power, low noise, and narrow linewidth lasers on TFLN chip. Here we achieve this goal, and demonstrate integrated high-power lasers on TFLN platform with up to 60~mW of optical power in the waveguides. We use this platform to realize a high-power transmitter consisting an electrically-pumped laser integrated with a 50~GHz modulator. 
\end{abstract}
\maketitle

Long haul telecommunication networks, data center optical interconnects, and microwave photonic systems rely on transmitting information using optical carriers \cite{winzer2018fiber,kikuchi2015fundamentals}. The ideal transmitter for these applications should operate over a large bandwidth with a small driving amplitude, emit high optical power, have negligible insertion loss, and be cost-effective. Recently, thin-film lithium niobate (TFLN) has emerged as a platform capable of achieving nearly all of these requirements \cite{zhu2021integrated}, using external lasers. The latter poses a major challenge to the achievable performance, complexity, and cost since it requires coupling between two discrete components. The ideal solution would integrate high-power and low-noise lasers on the TFLN photonic platform. Distributed feedback lasers (DFB) are excellent candidates for integration with TFLN because of their low cost, small footprint, and large output powers exceeding 100 mW. 
Such a solution would enable new architectures such as large arrays of high power transmitters as well as unprecedented performance in optical links
\cite{johansson2019advanced}. Here, we address this challenge by integrating distributed feedback (DFB) lasers, capable of producing of 170 mW - 200 mW of optical power, with TFLN integrated modulators featuring an electro-optic (EO) bandwidth in excess of 50 GHz \cite{wang2018integrated}. Using only passive-alignment and flip-chip thermo-compressive bonding, we integrate DFB lasers with pre-fabricated TFLN chips. With optimized overlap between the respective platform modes, we couple $\sim$ 60 mW of optical power into the TFLN waveguides.
\begin{figure}[ht!]
\centering
\fbox{\includegraphics[scale=0.95]{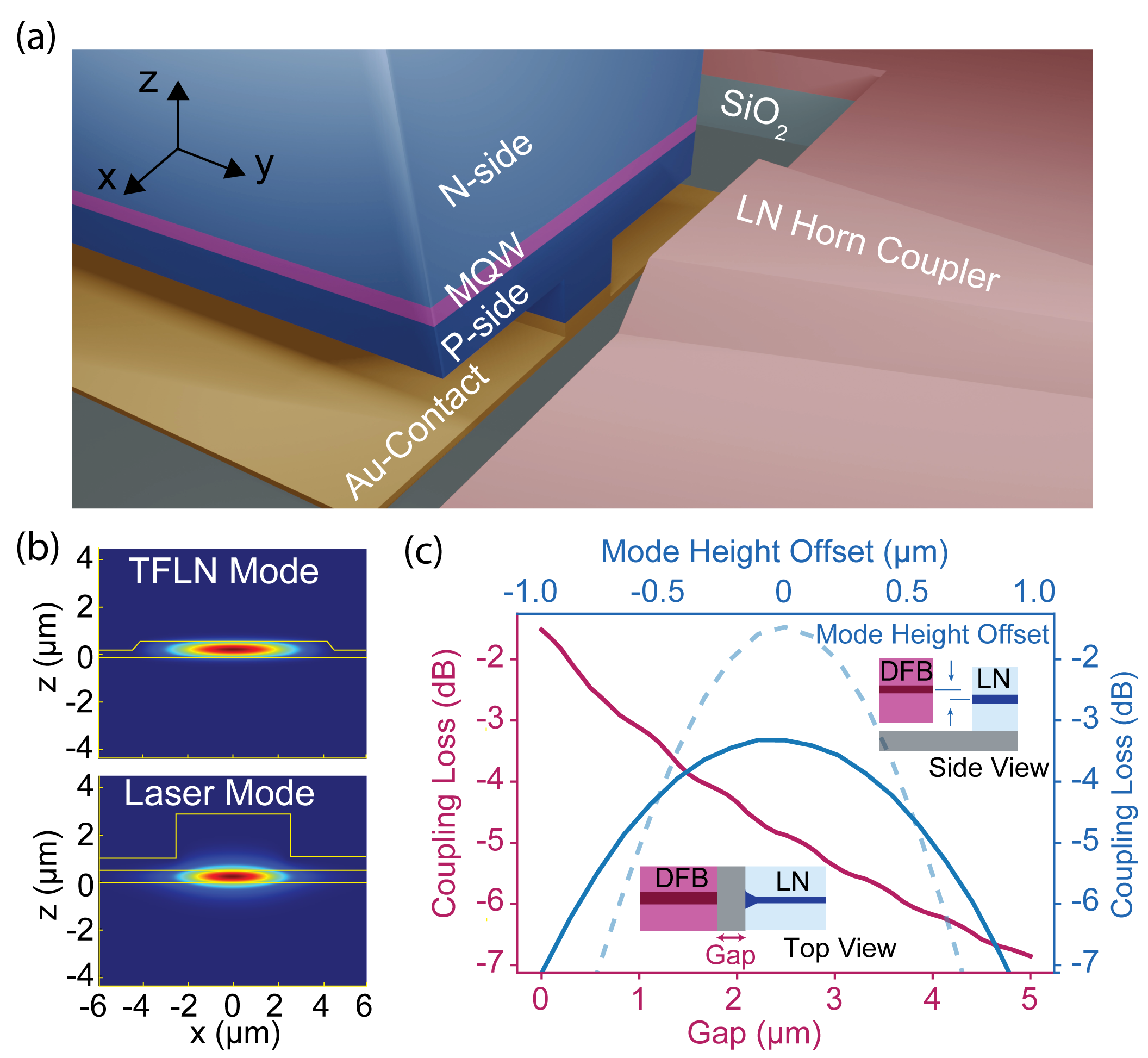}}
\caption{\textbf{Integration of Distributed Feedback Lasers (DFB) onto Thin-Film Lithium Niobate (TFLN):} \textbf{(a)} An illustration of the proposed approach, along with mode profiles of TFLN coupler and laser. DFB laser is flipped and the height is adjusted to ensure matched mode-heights between the two waveguides \textbf{(b)}. Optical mode profiles for both TFLN and DFB waveguides, obtained using a finite difference eigenmode solver. \textbf{(c)} Mode overlap simulations showing the coupling loss between the laser and TFLN waveguides as a function of the gap size (red) and height offset (blue). The latter are shown for gap size = 1.2$\mu$m (solid blue) and gap size = 0 $\mu$m (dashed blue ). Insets show the reference geometry for the gap and mode height offset calculation.}
\label{fig1}
\end{figure}

Our integration approach (Fig. 1a), relies on butt coupling between DFB and TFLN waveguides. 
This approach allows for an efficient injection of carriers through the active quantum well region, which is essential for envisioned high-power laser operation. First, TFLN devices are fabricated \cite{zhang2017monolithic,wang2018integrated} on a 600 nm thick, X-cut LN device layer that is bonded onto a 4.7 $\mu$m layer of thermally grown SiO\textsubscript{2} on top of Si substrate (NanoLN). The thickness of the buried oxide is chosen so that the optical mode-height of the flipped DFB laser waveguide ($\sim$ 4.66 $\mu$m) and TFLN waveguide ($\sim$ 5 $\mu$m) (Fig. 1b) are nearly matched. Mode height matching is later fine-tuned using gold deposition. The waveguide width is chosen to be 800 nm to ensure single mode operation. The waveguides are first defined using a negative-tone electron-beam resist (FOx-16, Dow Corning). The pattern is then transferred to TFLN by reactive ion etching using Ar\textsuperscript{+} to reach the slab height of 300-nm. We later deposit 800-nm of SiO\textsubscript{2} using plasma enhanced chemical vapor deposition as a device cladding. At the coupling region, the waveguides are tapered out to 8.2 $\mu$m in width. This horn coupler geometry ensures maximal overlap with the optical mode produced by the 5-$\mu$m wide DFB laser, and features high tolerances to lateral misalignment, mode height misalignment and separation (gap) between DFB and TFLN waveguides (Fig. 1c).
\begin{figure}[ht!]
\centering
\fbox{\includegraphics[scale=0.95]{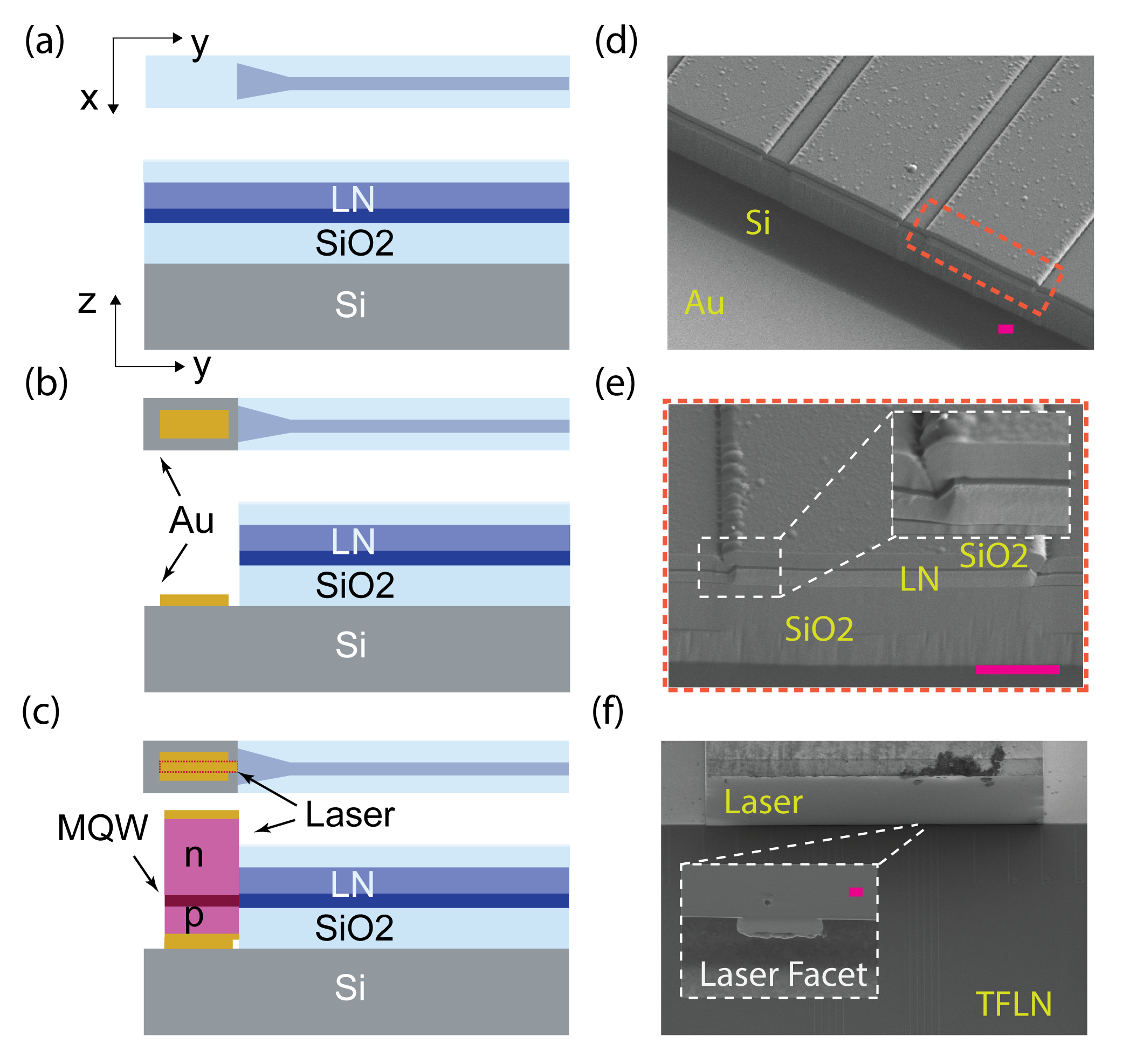}}
\caption{\textbf{Fabrication steps of the flip-chip bonding process} \textbf{(a)} Top-down and cross-section illustraton of the initial TFLN stack. \textbf{(b)} A trench is patterned and etched at the tip of the horn coupler. Then, a Ti/Pt/Au bonding pad (used for bonding and carrier injection) is defined 10-$\mu$m away from the edge of the groove using a second lithography step. \textbf{(c)} the DFB laser is flipped and bonded to the pad using thermo-compression Au-Au bonding (Fineplacer Femto 2). \textbf{(d)} Scanning electron microscope (SEM) image of the LN waveguide facets after trench fabrication. \textbf{(e)} A close-up SEM of the TFLN horn-coupler facet prior to bonding. \textbf{(f)} SEM image of the DFB laser bonded to TFLN platform. Inset shows the laser facet prior to bonding. Scale bars represent 1 $\mu$m.}
\label{fig2}
\end{figure}

The final step in TFLN chip fabrication is realization of a trench at the end of the horn coupler, where DFB lasers will be introduced. First, the upper SiO\textsubscript{2} cladding is etched using RIE with a mixture of C\textsubscript{3}F\textsubscript{8} and Ar gases. The LN layer is subsequently etched away using the same recipe as the waveguides,  followed by a buried oxide etch (Figs. 2(a-c)). 
Since the sidewall angle sets the minimum coupling gap between the DFB and TFLN waveguides, a smaller sidewall angle is desired to increase the coupling efficiency (Figs.2(d-e)). We evaluate the sidewall angle to be $\sim$85$^{\circ}$, which sets the minimum achievable gap between the laser and the waveguide to be $\sim$ 500 nm. Finally, a Ti/Pt/Au metal layer is deposited at the bottom of the trench using electron beam evaporation. This metallic layer serves as both one of the laser electrodes and the bonding pads. The laser is flip-chip bonded, P-side down, onto the metal layer at the bottom of the trench using gold-gold thermo-compression bonding (Fig. 2f). The mode height offset between the DFB and TFLN waveguides can be controlled by adjusting the thickness of the gold layer deposited in the trench.
\begin{figure}[ht!]
\centering
\fbox{\includegraphics[scale=0.95]{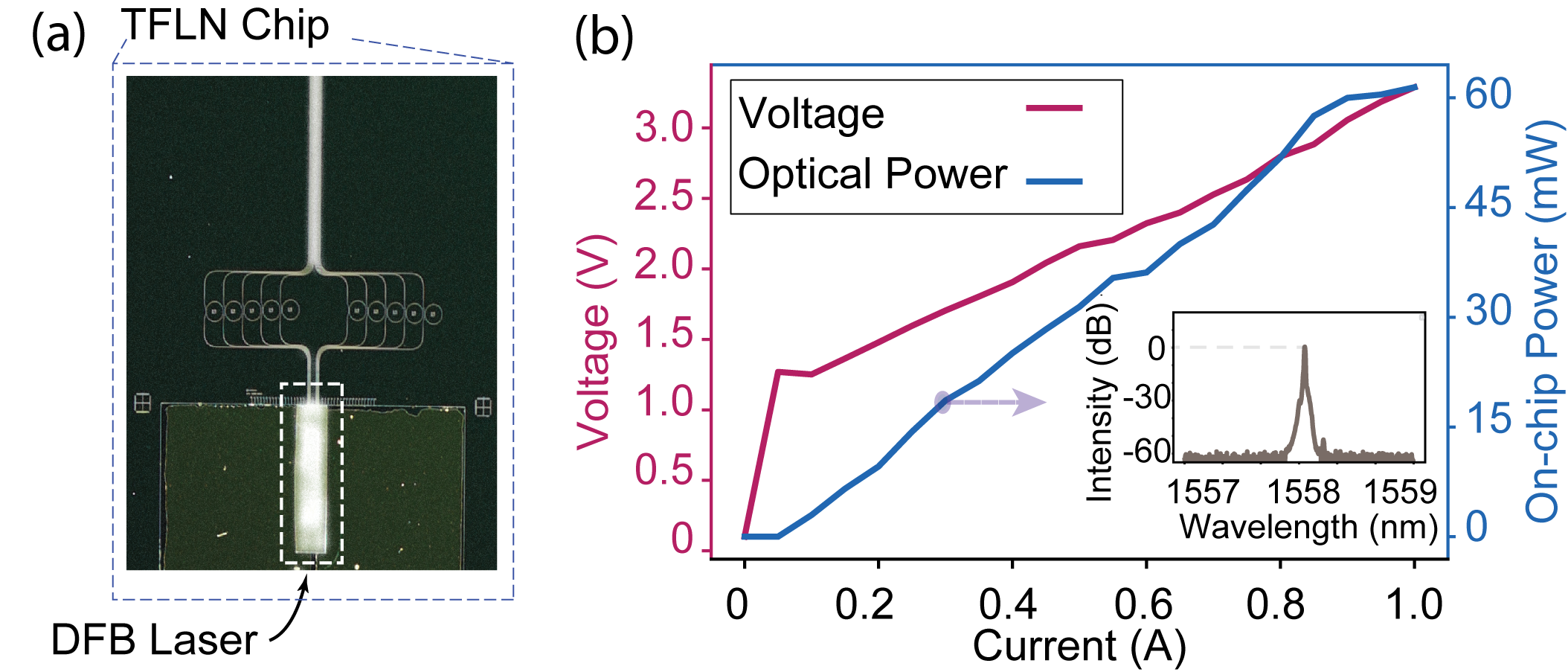}}
\caption{\textbf{Characterization of DFB laser integrated on passive TFLN device} \textbf{(a)} Microscope image of an exemplary passive device with bonded DFB laser. TFLN chip features several waveguides coupled to ring resonators. In this work, DFB emission wavelength is not in resonance with the rings.\textbf{(b)} Optical power inside TFLN waveguide and laser voltage as a function of the laser driving current. The inset shows the lasing spectrum confirming single mode operation. The emission wavelength of DFB is not in resonance with the ring resonator in this experiment.}
\label{fig3}
\end{figure}

On this passive device (Fig. 3a), the LIV measurements are performed by contacting a sourcemeter (Keithley 2400) to the N- and P-side of the laser and increasing the current up to 1.0 A (Fig. 3b). Clear evidence of single mode operation without mode hoping is observed for all integrated lasers tested (Fig. 3b - inset). The laser emission is collected from the TFLN device facet using a single lens with numerical aperture of 0.4. Taking into account the out-coupling losses, which we measure to be $4.8 \pm 0.5$ dB, we estimate an on-chip optical power in the range $60 \pm 7$ mW at 1.0 A, and under room temperature operation (no cooling). Such on-chip power is among the highest reported for all integrated photonics platforms \cite{theurer2019flip,huang2019high,9260967}. We achieve a coupling loss of $\sim$ 5 dB between the laser and TFLN waveguide which is due to sidewall slope, and alignment accuracy. Finally, the DFB linewidth is measured using a delayed self-heterodyne technique and found to be below 1 MHz. This is consistent with the linewidths measured on DFB lasers before integration.

To illustrate the full potential of our approach, DFB lasers are integrated with a TFLN EO-modulator (Fig. 4a). In this active device, a maximum of $\sim$ 25 mW (driving current $\sim$ 0.8 A) of optical power is estimated after the Mach-Zehnder interferometer (Fig. 4b). The lower power in this  active devices, compared to passive one, can be attributed to sub-optimal alignment as well as to  absorption loss introduced by metal electrodes in this modulator design that features small electrode gap  (electrode gaps$\sim$ 4 $\mu$m). The latter can be easily addressed by increasing the electrode separation, at the expense of reduced electro-optic efficiency.
\begin{figure}[ht!]
\centering
\fbox{\includegraphics[scale=0.95]{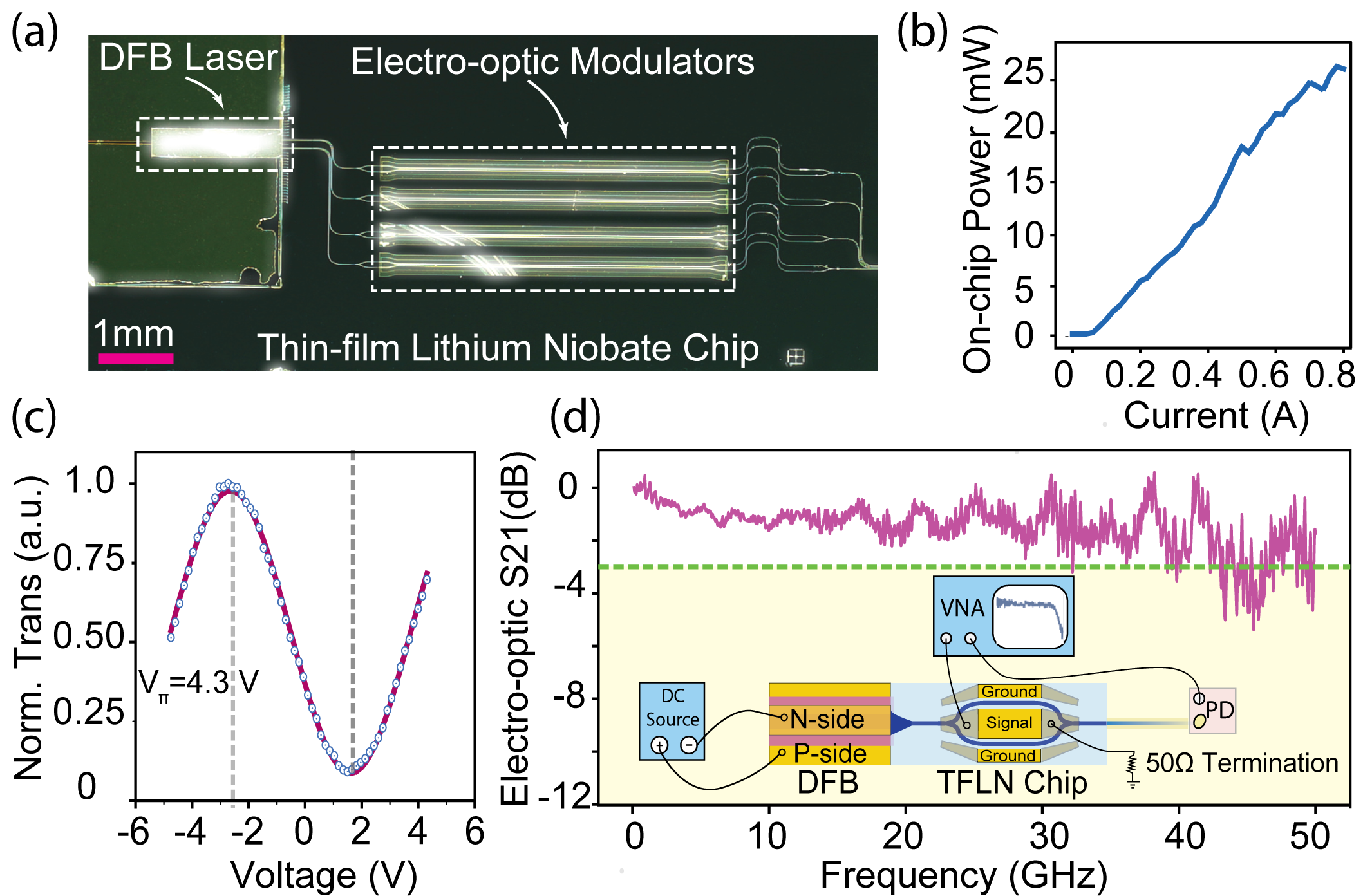}}
\caption{\textbf{
Electrically-pumped Integrated Laser-Modulator Transmitter
:} \textbf{(a)} Optical image of the transmitter, consisting of a DFB laser and EO intensity modulators.\textbf{(b)} Power in TFLN waveguide versus laser current for the device pictured in 2a. \textbf{(c)} Normalized optical transmission vs. the voltage applied to the modulator. Red line corresponds to fit, with V\textsubscript{$\pi$} = 4.3 V. \textbf{(d)} Small-signal electro-optic response of a device with an active modulation length of 5-mm and a 3-dB bandwidth of $\sim$ 50 GHz. S$_{21}$, transmission coefficient of the scattering matrix.}
\label{fig4}
\end{figure}

Electro-optic modulation is achieved using an integrated intensity modulator with 5-mm long segmented travelling wave electrodes \cite{kharel2021breaking}. The electrodes are contacted using ground-signal-ground probes (GSG) (GGB Model 50A) and terminated using a 50$\Omega$ termination. A measured $V_{\pi}$ of $4.3$ V demonstrates that the DFB emission can be modulated efficiently on-chip (Fig. 4c). To show the full-functionality of our platform as an integrated transmitter, we measure the high-frequency electro-optic response of our transmitter (S\textsubscript{21}) using a 50 GHz Vector Network Analyzer (VNA, Agilent), 45 GHz fast photo-diode (New Focus 1014), and a pair of 50 GHz GSG probes (Fig. 4d). RF cable losses, probes and detector response are subtracted from the measured frequency response. The response fluctuation beyond 45 GHz is due to the limited bandwidth of the photodiode.

In conclusion, we demonstrated the first high-power hybrid integrated transmitter on TFLN by flip-chip bonding a DFB laser. The bonding design allows efficient thermal anchoring for the laser which allowed uncooled operation with up to 60 (25) mW in the passive (active) waveguides. Our high-power transmitter platform will enable a new class of applications in digital and analog communication spaces. Beyond communication systems, integrating high-power lasers on other TFLN devices such as EO \cite{zhang2019broadband} and Kerr combs \cite{he2019self,Wang:18} enables a critical step towards realizing fully-integrated spectrometers \cite{shams2020integrated}. Additionally, other exciting frontiers include optical remote sensing and beam forming \cite{Morton:18}, photon pair generation \cite{zhao2020high}, and efficient frequency conversion \cite{Wang:18, lu2020toward} for realization of quantum networks. Future work would focus on realization of fully-integrated photonic links through integrating high-power and high-bandwidth photo-detector on TFLN platform \cite{7940062}. Recently, a complementary approach to TFLN-laser integration has been reported
\cite{OpdeBeeck:21}.
\newline
\newline
\textbf{Funding:} Defense Advanced Research Projects Agency (DARPALUMOS) (HR0011-20-C-0137). Air Force Office of Scientific Research (AFOSR) (FA9550-19-1-0376) 
\newline
\textbf{Acknowledgement:} We acknowledge assistance from Mehdi Rezaee, Tomas Stuopis, Dmitry Padrubny and Neil Sinclair. Device fabrication was performed at the Center for Nanoscale Systems (CNS) at Harvard University. 
\newline
\textbf{Disclosures:} L.H, M.Z, and M.L are involved in developing lithium niobate technologies at HyperLight Corporation.. H.G, and L.J are involved in developing DFB lasers at Freedom Photonics.
\newline
\textbf{Disclaimer:} The views, opinions and/or findings expressed are those of the author and should not be interpreted as representing the official views or policies of the Department of Defense or the U.S. Government.
\bibliographystyle{ieeetr}
\bibliography{laser}

\end{document}